\documentclass[preprint,prd,noshowpacs,nofootinbib]{revtex4}

\usepackage{color} 
\usepackage{amssymb}
\usepackage{amsmath}
\usepackage{graphicx} 
\usepackage{float}
\usepackage{braket}     
\usepackage{pdfpages}
\usepackage{epstopdf}
\usepackage[utf8]{inputenc}

\begin{document}


\begin{center}
{\huge{ Strings  versus Anti Strings in the inversion invariant or proper volume formulation }}  \\
\end{center}

\begin{center}
 E.I. Guendelman  \\
\end{center}

\begin{center}
\ Department of Physics, Ben-Gurion University of the Negev, Beer-Sheva, Israel \\
\end{center}

\begin{center}
\ Frankfurt Institute for Advanced Studies, Giersch Science Center, Campus Riedberg, Frankfurt am Main, Germany \\
\end{center}

\begin{center}
\ Bahamas Advanced Studies Institute and Conferences,  4A Ocean Heights, Hill View Circle, Stella Maris, Long Island, The Bahamas \\
\end{center}
E-mail:  guendel@bgu.ac.il,      \\

\date{\today}

\begin{abstract}
  The specific model studied is in the context of the modified measure formulation the string or branes where tension appear as an additional dynamical degree of freedom . We then consider the signed reparametrization invariant volume element formulation of dynamical strings and branes and find that the dynamical tension can produce positive tensions or negative tensions, corresponding exactly to strings and branes and anti strings and anti branes respectively. The anti strings are realized when a scalar time that defines the modified measure runs in the opposite direction to the world sheet time. For strings with positive tension, both times run in the same direction. The situation resembles the situation in Relativistic Quantum Mechanics with positive and negative energies, proper time of particles running forward with respect of coordinate time , while for anti particles proper tome runs opposite of coordinate time. An example where string anti string pair creation takes place in analogy to the  pair creation in an  external electric field in QED background field, this time in the presence of a background scalar field that couples to the strings and locally changes the tension, the tension field.
\end{abstract}

\maketitle

\section{Introduction}

String  Theories have been considered by many physicists for some time as the leading candidate for the theory everything,  including gravity, the explanation of all the known particles that we know and all of their known interactions (and probably more) \cite{stringtheory}. According to some, one unpleasant feature of string theory as usually formulated is that it has a dimension full parameter, in fact, its fundamental parameter , which is the tension of the string. This is when formulated the most familiar way.
The consideration of the string tension as a dynamical variable, using the modified measures formalism, which was  previously used for a certain class of modified gravity theories under the name of Two Measures Theories or Non Riemannian Measures Theories, see for example \cite{d,b, Hehl, GKatz, DE, MODDM, Cordero, Hidden}.  It is also interesting to mention that the modified measure approach has also been used to construct braneworld scenarios 
\cite{modified_measures_branes} and that Modified Measure Theories could be the effective Theories of Causal fermion systems \cite{MMandCFS}
Most recently, we have shown that modified measure gravity theories can be be formulated using an invariant volume element under Signed General Coordinate Invariance, that is for general coordinate transformations with negative jacobian  \cite{signedGR}, which could provide for a framework for the quantum creation of a baby universe as formulated by Farhi, Guth and Guven \cite{FARHIGUVENGUTH} which requires negatives measures for space time  and for Linde´s Universe multiplication scenario \cite{Lindemultiplication}, which also involves negative measures for space time in a parallel universe,
which requires the measure of integration to become negative at certain regions of the tunneling solution.

When applying these principles to string theory, this leads to the  modified measure approach to string theory, where  rather than to put the string tension by hand it appears dynamically.

This approach has been studied in various previous works  \cite{a,c,supermod, cnish, T1, T2, T3, cosmologyandwarped}. See also the treatment by Townsend and collaborators for dynamical string tension \cite{xx,xxx}, \cite{cosmologyandwarped}, which does not involve changing integration measures, so it cannot be used to achieve the goals presented in this paper however.

Here we will see that this modified volume element can play a similar role to the proper time, which for antiparticles runs in the opposite direction to laboratory time and how introducing appropriate background fields, string anti string pair creation could be possible.

\section{The Modified Measure String Theory }

The standard world sheet string sigma-model action using a world sheet metric is \cite{pol1}, \cite{pol2}, \cite{pol3},

\begin{equation}\label{eq:1}
S_{sigma-model} = -T\int d^2 \sigma \frac12 \sqrt{-\gamma} \gamma^{ab} \partial_a X^{\mu} \partial_b X^{\nu} g_{\mu \nu}.
\end{equation}

Here $\gamma^{ab}$ is the intrinsic Riemannian metric on the 2-dimensional string worldsheet and $\gamma = det(\gamma_{ab})$; $g_{\mu \nu}$ denotes the Riemannian metric on the embedding spacetime. $T$ is a string tension, a dimension full scale introduced into the theory by hand. \\

Now instead of using the measure $\sqrt{-\gamma}$ ,  on the 2-dimensional world-sheet, in the framework of this theory two additional world sheet scalar fields $\varphi^i (i=1,2)$ are considered. A new measure density is introduced:

\begin{equation}
\Phi(\varphi) = \frac12 \epsilon_{ij}\epsilon^{ab} \partial_a \varphi^i \partial_b \varphi^j.
\end{equation}

There are no limitations on employing any other measure of integration different than $\sqrt{-\gamma}$. The only restriction is that it must be a density under arbitrary diffeomorphisms (reparametrizations) on the underlying spacetime manifold. The modified-measure theory is an example of such a theory. \\

Then the modified bosonic string action is (as formulated first in \cite{a} and latter discussed and generalized also in \cite{c}

\begin{equation} \label{Action Mod Measure String1}
S = -\int d^2 \sigma \Phi(\varphi)(\frac12 \gamma^{ab} \partial_a X^{\mu} \partial_b X^{\nu} g_{\mu\nu} - \frac{\epsilon^{ab}}{2\sqrt{-\gamma}}F_{ab}(A)),
\end{equation}

where $F_{ab}$ is the field-strength  of an auxiliary Abelian gauge field $A_a$: $F_{ab} = \partial_a A_b - \partial_b A_a$. \\

It is important to notice that the action (\ref{eq:5}) is invariant under conformal transformations of the internal metric combined with a diffeomorphism of the measure fields, 

\begin{equation} \label{conformal}
\gamma_{ab} \rightarrow j \gamma_{ab}, 
\end{equation}

\begin{equation} \label{diffeo} 
\varphi^i \rightarrow \varphi^{'i}= \varphi^{'i}(\varphi^i)
\end{equation}
such that 
\begin{equation} \label{measure diffeo} 
\Phi \rightarrow \Phi^{'}= j \Phi
\end{equation}

Here $j$ is the jacobian of the diffeomorphim in the internal measure fields which can be an arbitrary function of the world sheet space time coordinates, so this can called indeed a local conformal symmetry.

To check that the new action is consistent with the sigma-model one, let us derive the equations of motion of the action (\ref{Action Mod Measure String1}). \\

The variation with respect to $\varphi^i$ leads to the following equations of motion:

\begin{equation} \label{eq variation of measure fields}
\epsilon^{ab} \partial_b \varphi^i \partial_a (\gamma^{cd} \partial_c X^{\mu} \partial_d X^{\nu} g_{\mu\nu} - \frac{\epsilon^{cd}}{\sqrt{-\gamma}}F_{cd}) = 0.
\end{equation}

since $det(\epsilon^{ab} \partial_b \varphi^i )= \Phi$, assuming a non degenerate case ($\Phi \neq 0$), we obtain, 

\begin{equation} \label{eq:a}
\gamma^{cd} \partial_c X^{\mu} \partial_d X^{\nu} g_{\mu\nu} - \frac{\epsilon^{cd}}{\sqrt{-\gamma}}F_{cd} = M = const.
\end{equation}

The equations of motion with respect to $\gamma^{ab}$ are

\begin{equation} \label{eq:8}
T_{ab} = \partial_a X^{\mu} \partial_b X^{\nu} g_{\mu\nu} - \frac12 \gamma_{ab} \frac{\epsilon^{cd}}{\sqrt{-\gamma}}F_{cd}=0.
\end{equation}

One can see that these equations are the same as in the sigma-model formulation . Taking the trace of (\ref{eq:8}) we get that $M = 0$. By solving $\frac{\epsilon^{cd}}{\sqrt{-\gamma}}F_{cd}$ from (\ref{eq:a}) (with $M = 0$) we obtain the standard string eqs. \\

The emergence of the string tension is obtained by varying the action with respect to $A_a$:

\begin{equation} \label{Tension Mod Measure String1}
\epsilon^{ab} \partial_b (\frac{\Phi(\varphi)}{\sqrt{-\gamma}}) = 0.
\end{equation}

Then by integrating and comparing it with the standard action it is seen that

\begin{equation}
\frac{\Phi(\varphi)}{\sqrt{-\gamma}} = T.
\end{equation}

That is how the string tension $T$ is derived as a world sheet constant of integration opposite to the standard equation (\ref{eq:1}) where the tension is put ad hoc.
Let us stress that the modified measure string theory action 
does not have any \textsl{ad hoc} fundamental scale parameters. associated with it. This can be generalized to incorporate super symmetry, see for example \cite{c}, \cite{cnish}, \cite{supermod} , \cite{T1}.
For other mechanisms for dynamical string tension generation from added string world sheet fields, see for example \cite{xx} and \cite{xxx}. However the fact that this string tension generation is a world sheet effect 
and not a universal uniform string tension generation effect for all strings has not been sufficiently emphasized before.

Notice that Each String  in its own world sheet determines its own  tension. Therefore the  tension is not universal for all strings.

\section{Elementary review of Particles and Antiparticles and Pair Creation in QED}
 From the 'Feynman' perspective, negative energy waves propagating into the past are physically realized as the antiparticles propagating into the future.  As he has shown \cite{Fyn:e-m} that from the classical equations of motion for a particle in an external field can be written as
\begin{equation} \label{LorentzForce1}
m\frac{d^{2}z^{\mu}}{d\tau^{2}}=e\frac{dz_{\nu}}{d\tau} F^{\mu \nu}
\end{equation} 
where $\tau$ is the proper time.  If we note as Feynman has that if we allow $\tau\rightarrow -\tau$ the equation becomes
\begin{equation}
m\frac{d^{2}z^{\mu}}{d\tau^{2}}=-e\frac{dz_{\nu}}{d\tau} F^{\mu \nu}
\end{equation} 
which is identical to the previous equation except that the particles charge has changed.  In other words, as far as its charge is concerned, it has become the antiparticle.  Thus, proper time running backward (i.e., $\tau \rightarrow -\tau $), while keeping the coordinate time unchanged.  led to the particle becoming an antiparticle.
Of course we can take the equivalent, but more suitable for our purposes transformation that we change the direction of coordinate time, while requiring that the proper time remains unchanged.

We are now in a position where we can discuss the scattering of a  particle in an external field.  Four possibilities are seen to exist: i) the scattering of the particle by the field, ii) the creation of a particle-antiparticle pair.  iii) the annihilation particle-antiparticle pair iv) the scattering of an antiparticle by the field. Below, we review one of them and in the following section we draw conclusions concerning the dynamical string theory and the analogy of the modified measure in the string case with the proper time in the particle case.
 \subsection{ Pair Creation in a Strong Uniform Electric Field in QED, importance of different directions of time and proper time for identifying particles and antiparticles}
 We shall present here a simplified version, by A.Vilenkin \cite{Vilenkin}, of the calculation of pair production first performed by Schwinger \cite{Schwinger1} 
 To begin our discussion, we consider a particle with charge $e$  and mass $m$ in a constant electric field.  The general equation of a particle in a field is most convenient written in terms of the Maxwell tensor $F_{\mu \nu}$ where for a constant electric $\mathbf{E}$ in the $x-$direction, $F^{01}=~-~E, \ \ F^{10}=E \ \ F^{0}_{\ 1}=E, \ \ F^{1}_{\ \ 0}=E$.  More explicitly, we have
 \begin{eqnarray}
 F^{\mu}_{\ \ \nu}=
 \left[ \begin{array}{llll} 
 0 & E & 0 &0 \\
 E & 0 & 0 & 0 \\
 0 & 0 & 0 & 0 \\
 0 & 0 & 0 & 0 
 \end{array} \right]
 \end{eqnarray}
The equation of motion for this  particle in a constant electric field is
\begin{equation}
m\frac{d^{2}x^{\mu}}{d\tau^{2}}=e F^{\mu}_{\ \ \nu}\frac{dx^{\nu}}{d\tau}
\end{equation}
\linebreak
The formal solution of for $u^{\mu}=\frac{dx^{\mu}}{d\tau}$ is $u^{\mu}(\tau)=\exp[\frac{e}{m}F^{\alpha}_{\ \ \beta}\tau]^{\mu}_{\nu}u^{\nu}(0)$
The exponential can be expanded and we have
\begin{eqnarray} \label{expo}
\exp[\frac{e}{m}F^{\alpha}_{\ \ \beta}\tau]^{\mu}_{\nu}=\delta^{\mu}_{\nu}+\frac{e}{m}\tau E \Delta^{\mu}_{\nu}+\frac{1}{2}(\frac{e}{m}\tau E \Delta^{\mu}_{\nu})^{2}+\cdots
\end{eqnarray}
where
\begin{eqnarray}
\Delta=
\left[\begin{array}{llll}
0 & 1 & 0 & 0 \\
1 & 0 & 0 & 0 \\
0 & 0 & 0 & 0 \\
0 & 0 & 0 & 0 
\end{array} \right]
\end{eqnarray}
Separating even and odd power in Eq. \ref{expo} we have
\begin{eqnarray}
u^{0}&=&\cosh\left(\frac{eE\tau}{m}\right)u^{0}(0)+\sinh\left(\frac{eE\tau}{m}\right)u^{1}(0)  \nonumber \\
u^{1}&=&\sinh\left(\frac{eE\tau}{m}\right)u^{0}(0)+\cosh\left(\frac{eE\tau}{m}\right)u^{1}(0)
\end{eqnarray}

$\\$
Integrating with respect to $\tau$ yields (where we have dropped arbitrary constants of integration)
\begin{eqnarray}
x^{0}&=&\frac{m}{eE}\left\{\sinh\left(\frac{eE\tau}{m}\right)u^{0}(0)+\cosh\left(\frac{eE\tau}{m}\right)u^{1}(0)\right\} \nonumber \\
x^{1}&=&\frac{m}{eE}\left\{\cosh\left(\frac{eE\tau}{m}\right)u^{0}(0)+\sinh\left(\frac{eE\tau}{m}\right)u^{1}(0)\right\}
\end{eqnarray}

We choose the following boundary conditions $u^{1}(0)=0,\ \ u^{0}(0)=1$ which leads to
\begin{eqnarray} \label{particle}
x^{0}&=&\frac{m}{eE}\sinh\left(\frac{eE\tau}{m}\right) \nonumber \\
x^{1}&=&\frac{m}{eE}\cosh\left(\frac{eE\tau}{m}\right)
\end{eqnarray}
and for the boundary condition  $u^{1}(0)=0,\ \ u^{0}(0)=-1$  leads to
\begin{eqnarray} \label{antiparticle}
x^{0}&=&-\frac{m}{eE}\sinh\left(\frac{eE\tau}{m}\right) \nonumber \\
x^{1}&=&-\frac{m}{eE}\cosh\left(\frac{eE\tau}{m}\right)
\end{eqnarray}
The solution given by Eq. \ref{particle} represents a particle solution while the solution of Eq. \ref{antiparticle} represents the anti-particle solution. Both solutions, together satisfy

\begin{equation} \label{hyperbola} (x^{1})^{2}-(x^{0})^{2}=\left(\frac{m}{eE}\right)^{2}\end{equation}

At classical level, these solutions are distinct and one solution can not evolved into the other.  Thus, a particle couldn't evolve to an anti-particle.  However, the semi-classical approximation which consists of considering the classical equations of motion but with imaginary time.  Then inserting $t=-it_{E}$
we obtain that the hyperbola of Eq. \ref{hyperbola} becomes a circle
\begin{equation}
(x^{1})^{2}+t_{E}^{2}=\left(\frac{m}{eE}\right)^{2}
\end{equation}
This tunneling solution can now interpolate between the anti-particle and particle solutions.  (see Fig. 2).
In the imaginary time region, the action, $S=-iS_{E}$ where $S_{E}$ is given by 
\begin{equation}
S_{E}=\int dt_{E} \left\{ m\sqrt{1+\left(\frac{dx}{dt_{E}}\right)^{2}}-eEx\right\}
\end{equation}
which is the analytic continuation of Eq. \ref{action} and Eq. \ref{actionE&M}.
Introduction the angular variable $\theta$ where $x=\frac{m}{eE}\cos \theta,\ \ t_{E}=\frac{m}{eE}\sin \theta$ we obtain that $S_{E}=\pi m^{2}/eE$.  Since the probability is given (up to pre factors) by $\exp{(-S_{E})}=\exp(-\frac{\pi m^{2}}{eE})$ for $eE>0$.
Notice that the distance of the particle and antiparticle at the moment of creation is $\Delta x=\frac{m}{eE}$ which has a physical interpretation is manifest in writing it as 

\begin{equation}
W=eE\Delta x=2mc^{2}
\end{equation}
$\\$
where we have restored the 'c' to make its physical meaning clearer.  Thus we can create a pair of particle-antiparticle in a constant electric field by performing work, $W$ in a distance $\Delta x$ equal to the the sum of the rest masses of the particles, i.e. $2mc^{2}$. In the case for an electric field that doesn't extend through all of space, still it must be extended enough to perform the work equal to the sum rest masses of the two particles in order to create a pair.

\section{Extension to signed reparametrization invariant Modified Measure volume element as the proper volume of Strings}
In  \cite{signedGR}, we discussed the generalization of general coordinate transformations to signed general coordinate transformations and found the modified volume which is invariant under signed general coordinate transformations.

For the case of strings now notice that under a signed general reparametrization transformation of the  world sheet, that could for example change the direction of time, 
$$ d^2 \sigma  \rightarrow J d^2 \sigma $$ , while 
$$ \sqrt{ -\gamma}  \rightarrow  \mid J \mid ^{-1}\sqrt{-\gamma} $$
where $J$ is the jacobian of the transformation and $ \mid J \mid$ is the absolute value of the jacobian of the transformation. Therefore 

$$d^2 \sigma {\sqrt{-\gamma}} \rightarrow 
\frac{J}{ \mid J \mid} d^2 \sigma  {\sqrt{-\gamma}}$$ , 

so invariance of the volume element is achieved only for $J = \mid J \mid$, that is if $J>0$, that is signed general coordinate transformations are excluded if we want to define signed invariant reparametrization volume element. 

Notice that we are not discussing invariance of the  action here, which is a different matter,  
since under a signed reparametrization transformation the time direction may change for example.
Then the limits of integration change and when restoring back the limits of integration we obtain anothe minus sign and the action restores its invariance.

However what we want to know is, if we now can have a measure that does not change sign, even if we change for example the direction of the coordinate time, which will be the corresponding generalization  of the proper time, which continues to increase as the particle goes backwards in time and now in the case of the strings, the modified volume can be monotonically increasing even as the strings go backwards in time, Furthermore, this will imply that there are effects that
cannot be described in the standard formalism, like string anti string pair creation, which would require the generalization  of the proper time, which continues to increase as the string goes backwards in time . 

Indeed , replacing measure $\sqrt{-\gamma}$ by the modified measure $\Phi $ provides us with such measure since $\Phi $  transforms according to $ J ^{-1}$ instead of $ \mid J \mid ^{-1}$. So we can have now a positive defined measure for strings propagating backwards in the string sheet time in exact analogy to the QED case.

Notice that $$\Phi(\varphi) d^2 \sigma = d\varphi_1 d\varphi_2 $$

So the integration domain in the $\varphi_1 \varphi_2 $ space is invariant under these signed reparametrizations, since 
$\varphi_1 \varphi_2 $  are both scalars and do not transform under inversions   .
So we can take the time of the string, say $\sigma_1$ to go in the opposite direction to that of the coordinate time, thus defining an anti string, the same way as the antiparticle was defined when the proper time and the coordinate time ran in different directions. $\varphi_2 $ could be taken as the spacial coordinate of the string, periodic for a closed sting for example, and will not play a role in the discussion on the difference between strings and anti strings.  Generically, the consideration of the $\varphi_1 \varphi_2 $ space
as opposed to the $\sigma_1 \sigma_2 $ space enlarges the space of possible configurations and solutions and allows for example the consideration of pair creation or pair annihilation , which requires non trial mappings between the  $\sigma_1 \sigma_2 $ space and the  $\sigma_1 \sigma_2 $ spaces .
Furthermore if we want the full action density to be invariant under signed reparametrizations,  in (\ref{Action Mod Measure String1)}, we have to correct the term $$\Phi(\varphi) \frac{\epsilon^{ab}}{2\sqrt{-\gamma}}F_{ab}(A)$$, since $\Phi(\varphi)$ and
$\sqrt{-\gamma}$ differ by a sign under signed reparametrization transformation, To cure this, we replace this by 
$$\Phi(\varphi)^2 \frac{\epsilon^{ab}}{2(-\gamma)}F_{ab}(A)$$,

obtaining the action invariant under signed reparametrization invariant, Where both the domain of integration and the lagrangian density are invariant under signed reparametrizations transformations.

\begin{equation} \label{Action signed Mod Measure String1}
S = -\int d^2 \sigma \Phi(\varphi)(\frac12 \gamma^{ab} \partial_a X^{\mu} \partial_b X^{\nu} g_{\mu\nu} - \Phi(\varphi)\frac{\epsilon^{ab}}{2(-\gamma)}F_{ab}(A)),
\end{equation}
now the equation of motion that comes from the variation of the internal gauge field is ,
\begin{equation} \label{Tensions in signedMod Measure String1}
\epsilon^{ab} \partial_b (\frac{\Phi^2(\varphi)}{(-\gamma))}) = 0.
\end{equation}

which is integrated to,
\begin{equation}
\frac{\Phi^2(\varphi)}{(-\gamma)} = T^2
\end{equation}
where without loosing generality we can take $T$ positive. The above equation implies that positive tensions are accompanied with negative ones,
\begin{equation}
\frac{\Phi(\varphi)}{\sqrt{-\gamma}} = \pm T.
\end{equation}
the negative tension strings we associate with the ¨anti-strings¨. We can see that the results obtained from the signed reparametrization invariant action are  (\ref{Action signed Mod Measure String1}) to those obtained in the previous section, except for the possibility that there are positive and negative tensions. To start with, the equation that replaces 
(\ref{eq variation of measure fields}) is
\begin{equation} \label{eq:variantion of measure fields in signed}
\epsilon^{ab} \partial_b \varphi^i \partial_a (-\gamma^{cd} \partial_c X^{\mu} \partial_d X^{\nu} g_{\mu\nu} + 2 \Phi \frac{\epsilon^{cd}}{(-\gamma)}F_{cd}) = 0.
\end{equation}
which again can be integrated, assuming non singular measure $\Phi$ , 
\begin{equation} \label{integration of eq:variantion of measure fields in signed}
-\gamma^{cd} \partial_c X^{\mu} \partial_d X^{\nu} g_{\mu\nu} +   \frac{\Phi\epsilon^{cd}}{(-\gamma)}F_{cd} = M.
\end{equation}
The variation with respect to $\gamma^{ab}$ gives now, for $\Phi \neq 0$ as
\begin{equation} \label{new T}
T_{ab} = \partial_a X^{\mu} \partial_b X^{\nu} g_{\mu\nu} - \Phi \gamma_{ab} \frac{\epsilon^{cd}}{(-\gamma)}F_{cd}=0.
\end{equation}
Taking the trace of (\ref{integration of eq:variantion of measure fields in signed}) and comparing with ( \ref{new T}) we obtain $M=0$ then
 solving $\frac{\Phi\epsilon^{cd}}{(-\gamma)}F_{cd}$, from \ref{integration of eq:variantion of measure fields in signed} now for 
$M=0$ and inserting in (\ref{new T}) we recover the standard Polyakov equation for a string, but now we have two species of strings, with positive and negative tensions.

Notice however that from a ¨practical¨ point of view,  it is not necessary to work with an action density that is completely invariant under the signed reparametrizations, in particular, we can still work with  the original action \ref{Action Mod Measure String1}, which avoids quadratic equations for the tension, but still allows for positive and negative tensions, just as in the modified version that has complete signed reparametrization invariance and is simple to manipulate, since everything becomes linear. We now consider possible ways to scatter strings from an external field so that the tension may change, still in the most simple formalism.
\section*{Introducing a New Background Field, The Tension Field} 
Schwinger \cite{Julian Schwinger} had an important insight and understood that all the information concerning a field theory can be studied by understanding how it  reacts to sources of different types. Different types of background bulk fields, like the dilaton field, the two index antisymmetric gauge field have been discussed in the text book by Polchinski for example  \cite{Polchinski} . 
Here , instead of the traditional background fields usually considered in conventional string theory, one may consider another scalar field that induces currents in the string world sheet and since the current couples to the world sheet gauge fields, this produces a dynamical tension controlled by the external scalar field as shown at the classical level in \cite{Ansoldi}. In the next two subsections we will study how this comes about in two steps, first we introduce world sheet currents that couple to the internal gauge fields in Strings and Branes and second we consider a coupling to an external scalar field by defining a world sheet current  that is induced by such external scalar field.and then  coupling this current to the internal gauge fields of the Strings .
After this is done, we argue that such external field could lead to string anti string pair creation, in analogy to the QED process of electron anti electron pair production. So the Tension Field probes exactly the string - anti string structure of the modified measure string theory.

\subsection*{Introducing world sheet currents that couple to the internal gauge fields and locally change the tension}

If to the action of the string  we add a coupling
to a world-sheet current $j ^{a}$,  i.e. a term
\begin{equation}
    S _{\mathrm{current}}
    =
    \int d ^{2} \sigma
        A _{a}
        j ^{a}
    ,
\label{eq:bracuract}
\end{equation}
 then the variation of the total action with respect to $A _{a }$
gives
\begin{equation}
    \epsilon ^{a b}
    \partial _{a }
    \left(
        \frac{\Phi}{\sqrt{- \gamma}}
    \right)
    =
    j ^{b}
    .
\label{eq:gauvarbracurmodtotact}
\end{equation}
We thus see indeed that, in this case, the dynamical character of the
brane is crucial here.
\subsection*{How a world sheet current can naturally be induced by a bulk scalar field, the Tension Field}

Suppose that we have an external scalar field $\phi (x ^{\mu})$
defined in the bulk. From this field we can define the induced
conserved world-sheet current
\begin{equation}
    j ^{b}
    =
    e \partial _{\mu} \phi
    \frac{\partial X ^{\mu}}{\partial \sigma ^{a}}
    \epsilon ^{a b}
    \equiv
    e \partial _{a} \phi
    \epsilon ^{a b}
    ,
\label{eq:curfroscafie}
\end{equation}
where $e$ is some coupling constant. The interaction of this current with the world sheet gauge field  is also invariant under local gauge transformations in the world sheet of the gauge fields
 $A _{a} \rightarrow A _{a} + \partial_{a}\lambda $.

For this case,  (\ref{eq:gauvarbracurmodtotact}) can be integrated to obtain
\begin{equation}
  T =  \frac{\Phi}{\sqrt{- \gamma}}
    =
    e \phi + T _{0}
    ,
\label{eq:solgauvarbracurmodtotact2}
\end{equation}

The constant of integration $T _{0}$ could represent the initial value of the tension before the string enters the region where the tension field is present. Notice that the interaction is metric independent since the internal gauge field does not transform under the the conformal transformations. This interaction does not therefore spoil the world sheet conformal transformation invariance in the case the field $\phi$ does not transform under this transformation.  Therefore, one way to dynamically change the string tension tension dynamically is through the introduction this new background field, that changes the value of the the tension locally in the world sheet \cite{Ansoldi} . This background field was then called the  ¨tension field¨ 
and used for many different scenarios \cite{cosmologyandwarped}, \cite{Escaping}, the results of which have been summarized in \cite{summary}  and most recently for the construction of braneworld scenarios in dynamical string tension theories \cite{Life} and
\cite{Lidhtlikeandbraneworld}, and as we will discuss in the next section, this tension field could provide a way to communicate between strings and anti strings.
\section{Scattering of a dynamical tension string field leading to string anti string pair creation in analogy with the analog QED process}
The QED example explained before in this paper inspires an analogous model for a possible scenario for a string-anti string creation process. Here instead of the the external electric field we can have the tension field. 

We  specify the initial state according to the $\varphi_1$ time as the anti string, where the time
$\varphi_1$ and the world sheet coordinate time $\sigma_1 $ (running in the same direction as the coordinate time $t$) run in opposite directions, so the tension is therefore negative. Assume $T _{0}$ that  could represent the initial value of the tension before the string enters the region where the tension field is present is negative, 

Then as the $\varphi_1$ time continues to increase, and the world sheet time continues to decrease, 
we encounter a region where the tension field $\phi$ is present , and over this region 
the tension field suffers a a positive jump $e\Delta \phi$ which can be bigger than $-T_{0}$, turning the tension from negative to positive. This is the analog of the condition for pair creation in the QED case, that is that the electric field is extended id a region of space with minimum extension given by $W=eE \Delta x=2mc^{2}$.
This again,  as in the QED case, cannot happen classically, there should be a tunneling region that connects the anti string solution going backwards in coordinate time with the string solution going forward in coordinate time.   Unfortunately we cannot yet present full details as compared with the QED case, because on the much higher complexity of the string dynamics as compared with the particle dynamics of the QED problem, but we plan to present more details in a future publication.

In the case we were to use the full signed reparametrization formulation the equation \ref{eq:solgauvarbracurmodtotact2} has to be modified and if the coupling to the tension field is the same, now becomes,
\begin{equation}
  T^{2} =  \frac{\Phi^{2}}{(- \gamma)}
    =
    e \phi + T _{0}
    ,
\label{eq:solgauvarbracurmodtotactfromsignedinv}
\end{equation}
As we see, there is a difference with the simpler case studied before, since the above equation involves the square of the dynamical tension, so there must be a boundary for the evolution of the tension field giver by $ e \phi + T _{0}=0$ and  after this point the tension field must bounce in the opposite direction, and if particle antiparticle is desired, when bouncing in the opposite direction the opposite sign of the tension must be taken. If the motion of the tension field is symmetric around the bounce, the string and anti string will have exactly opposite tensions, something that is not necessary in the simplified model which is not signed reparametrization invariant.

\section{Discussion on  Signed reparametrization  invariant volume element for  Brane Theories. Analogy with Relativistic QM positive and negative energies.}
The $d$ dimensional extended object ($d=1$ corresponds to the string case) has an action that provides for a dynamically generated brane tension  

\begin{equation} \label{Action Mod Measure Brane1 Theoy}
S = -\int d^{d+1}\sigma \Phi(\varphi)(\frac12 \gamma^{ab} \partial_a X^{\mu} \partial_b X^{\nu} g_{\mu\nu} - \frac{\epsilon^{a_1 a_2.....a_{d+1}}}{2\sqrt{-\gamma}}F_{a_1 a_2.....a_{d+1}}(A)),
\end{equation}
where $F_{a_1 a_2.....a_{d+1}}(A)$ is a field stregth deriving from the totally anti symmetric derivative of a totally antisymmetric potential with $d$ indices. Also the measure is constructed from the derivatives of $d+1$ scalars.
The variation with respect to those totally antisymmetric potential with $d$ indices gives the equation that the derivative of $\frac{\Phi(\varphi)}{\sqrt{-\gamma}}$ is a constant $T$, which we identify with the tension. 
(\ref{Action Mod Measure Brane1 Theoy}) is not however invariant under signed reparametrization transformations, in order to do this $\sqrt{-\gamma}$  sould not appear, only $-\gamma$, so the c signed invariant generalization would be, 
\begin{equation} \label{signed Action Mod Measure Brane1 Theoy}
S = -\int d^{d+1}\sigma \Phi(\varphi)(\frac12 \gamma^{ab} \partial_a X^{\mu} \partial_b X^{\nu} g_{\mu\nu} - \frac{ \Phi(\varphi) \epsilon^{a_1 a_2.....a_{d+1}}}{2(-\gamma)}F_{a_1 a_2.....a_{d+1}}(A)),
\end{equation}
from the variation of the internal gauge fields and integration of the resulting equation,  we get now $\frac{\Phi^2(\varphi)}{(-\gamma)} = T^2$, just as in the string case, which again leads to positive and negative tensions, i.e. $\frac{\Phi(\varphi)}{\sqrt{-\gamma}} = \pm T$. 
As in the modified measure string, it is not strictly necessary to work with \ref{signed Action Mod Measure Brane1 Theoy},  \ref{Action Mod Measure Brane1 Theoy} also works.

As in relativistic quantum mechanics \cite{RLQM} one may attempt to eliminate negative solutions, which in the case of relativistic quantum mechanics are negative energy solutions, that attempt, known as the Relativistic Schroedinger equation fails  (\cite{RLQM}) and such equation cannot even describe the scattering by a square well potential. As pointed out by Feynman, this is related to the fact that the space of positive energy solutions is not large enough to represent an arbitrary initial condition. The correct interpretation of the Klein paradox \cite{RLQM} must involve particles and antiparticles, the alternative of eliminating the negative energy solutions does not work. 

We introduced a new background field,the tension field, which is specially designed to probe the string anti string dynamics of the modified measure string theories. The process of string anti string creation can be studied qualitatively for a simpler formulation that is not totally invariant under signed reparametrizations. in this case the resulting pair  of string anti and string may not have exactly opposite tensions. In the signed reparametrization invariant formulation, we
find that zero tension squared is a special point from which it must bounce back and transition from  negative to positive values is possible, in most likelihood this bounce could be symmetric, leading to pair creation with exactly opposite tensions for the string and the anti string.

Concerning simultaneous use of positive and negative brane tensions, we can point out that this has been used extensively in brane world models, see for example \cite{RS}. The interpretation of the negative tension branes as antibranes could be useful. 

One interesting aspect of the dynamical strings and brane, in that due to the influence of the tension field a part of the string or brane could be anti string or anti brane while the rest could be string or brane, i.e., have positive tension. One may have an anti brane bubble in the midst of a brane environment, since the transition for brane to anti brane  or the reverse due to the tension field is local.

Finally, the tension field that locally changes the value of the string tension has been introduced as a background field, so far just an external field. To have a complete picture, more
work should be done to determine it dynamically Some ideas in this directions were discussed for a system containing two types of strings with different integration constants  $T_{1}$ and $T_{2}$
and imposing separately conformal invariance on the two string types separately, which lead to an algebraic equation for the  tension field, that indicates the formation of a braneworld at large times \cite{Life}, \cite{Lidhtlikeandbraneworld}. An interesting possibility could be if those two types of strings could be strings and anti strings generated by a pair creation process like the one that we have discussed here, this may complete the picture described in those paper, including the quantum formation of such braneworlds.

A different approach to dynamical string tension and the possibility of negative string tension has been discussed in \cite{Paul}, the relation of the approach of this paper is not straightforward, but could exist at the level of potentially similar effects.

\section{Acknowledgments}
  I  want to thank Stefano Ansoldi  Emil Nissimov, Svetlana Pacheva and Paul Steinhardt for conversations on dynamical string tensions, the Foundational Questions Institute (FQXi)  and the COST actions  Quantum Gravity Phenomenology in the multi messenger approach, CA18108 , COST  Gravitational waves, Black Holes and Fundamental Physics, CA16104 and COST COSMOVERSE,  Action CA21136 for financial support.


\end{document}